\documentclass[letterpaper]{amsart}
\pdfoutput=1

\theoremstyle{definition}

\theoremstyle{remark}

\usepackage{hyperref}
\usepackage{graphicx}
\usepackage{color}
\usepackage{amsmath}
\usepackage{amsfonts}
\usepackage{amssymb}
\usepackage{epsfig}
\usepackage{rotating}
\usepackage{graphics}

\def\beq{\begin{equation}}
\def\eeq{\end{equation}}
\def\bea{\begin{eqnarray}}
\def\eea{\end{eqnarray}}
\def\ar{\begin{array}}
\def\ear{\end{array}}

\def\nn{\nonumber}

\def\ga{\gamma}

\def\si{\sigma}

\def\de{\delta}

\def\eps{\epsilon}

\begin{document}
\begin{center}
{\Large Structures in the Planck map of the CMB}\\[5mm]

Daniel An$^1$, Krzysztof A. Meissner$^{2}$, Pawe\l~ Nurowski$^{3}$\\[3mm]
{\it $^1$Science Department SUNY Maritime College, 6 Pennyfield Av., New York  10465, USA\\
$^2$ Faculty of Physics, University of Warsaw, Ho\.za 69, 00-681 Warsaw, Poland\\
$^3$ Center for Theoretical Physics of PAS, Al. Lotnik{\'o}w 32/46, 02-688 Warsaw, Poland
}

\begin{abstract}
\noindent We present the results of the quest for ring-type structures on the maps observed by the Planck satellite.
\end{abstract}
\end{center}

\vspace{0.2cm}

\noindent
In this brief note we announce the results of a preliminary quest for the ring-type structures on the CMB maps observed by the Planck collaboration \cite{PC}. This work corroborates our earlier analysis on WMAP data \cite{MNR} which was inspired by predictions of Roger Penrose \cite{P}.

The present paper reports on analysis of the real CMB temperature map in the
frequency band 70 GHz as measured by the Planck collaboration. To manipulate the maps we used the HEALPix code \cite{HE}. 300 simulated maps (all with $k=10$) were created and analyzed in the same way for a comparison with the real map.

The procedure applied to look for the ring-type structures consisted in the following.

\begin{enumerate}

\item
a grid of points with HEALPix $k=4$ parametrization spreading over the entire
celestial sphere has been created
\item
we have excluded from all the
maps the Milky Way belt, $\pm 0.4$ radians above and below the Galactic equator 
\item
We were considering circles $C_{(i,\ga)}$ with a spherical radius $\gamma$ centered at the $N_d=960$ points $(\theta^i,\phi^i)$, $i=1,2,\dots,N_d$, from the grid on the sphere. Each circle $C_{(i,\ga)}$ was surrounded by two rings - an inner ring $R_{(i,\ga,\eps_-)}$, and an outer ring $R_{(i,\ga,\eps_+)}$ - each of width $\eps$. The inner (respectively, outer) ring consisted of $N_{(i,\ga,\eps_-)}$ (respectively, $N_{(i,\ga,\eps_+)}$) points, whose spherical angle from the point $(\theta^i,\phi^i)$ was between $\ga-\eps$ and $\ga$ (respectively, between $\ga$ and $\ga+\eps$). The points in the rings where taken from the HEALPix grid with $k=10$. 
\item For each $(\theta^i,\phi^i)$, each $\ga=0.02$, 0.04, 0.08, 0.12, 0.16, 0.20, 0.24, 0.28, 0.3, 0.32, 0.34 radians, and each $\eps=0.01$, 0.02, 0.04, 0,08 radians, we calculated the difference between the average temperature in the inner ring and outer ring, i.e. the quantity 
\beq
I_{(\ga,\eps)}(\theta^i,\phi^i)=\sum_{{\rm points~in}~R_{(i,\ga,\eps_-)}}\hspace{-0.5cm}\frac{\de T_m}{N_{(i,\ga,\eps_-)}}\quad-\sum_{{\rm points~in}~R_{(i,\ga,\eps_+)}}\hspace{-0.5cm}\frac{\de T_m}{N_{(i,\ga,\eps_+)}}\nn\quad
\eeq
with $\de T_m$ being the temperatures at the points in the respective rings as 
given by the Planck data team. 
\item We were looking for the points $(\theta^i,\phi^i,\ga,\eps)$ for which $I_{(\ga,\eps)}(\theta^i,\phi^i)$ had large absolute values.
\end{enumerate}

\vspace{0.1cm}
\noindent{\bf 4. Results}

It is commonly taken for granted (with the notable exception of \cite{GP}) that the temperature distribution in the Cosmic Microwave Background is purely statistical being produced by the quantum fluctuations usually assumed to have taken place during inflation (as evolution of quantum fields in De Sitter space suggests).

\begin{figure}[htb]
\begin{center}
\includegraphics[scale=0.25]{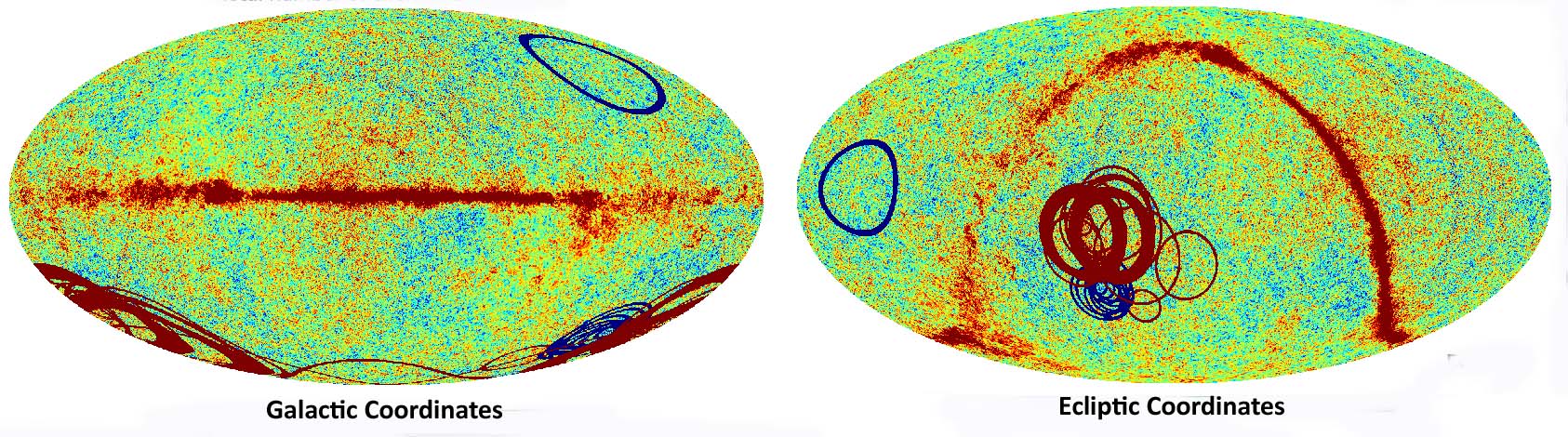}
\caption{\scriptsize{Plot of rings of width $\epsilon=0.08$ radians present on the Planck CMB temperature map in the frequency band 70 GHz. Observe a large concentration of such rings on the Southern Galactic hemisphere centered at $(\tilde\theta_1,\tilde\phi_1)=(2.60,3.70)$ with blue rings, and at $(\tilde\theta_2,\tilde\phi_2)=(2.59,2.89)$ with red rings. Red and blue rings are rings that have significant positive or negative $I_{(\ga,\eps)}$ values respectively.}}
\end{center}
\end{figure}

Our comparison between the distributions of large values of quantities $I_{(\ga,\eps)}(\theta^i,\phi^i)$ for the real Planck temperature map and purely statistical HEALPix maps may weaken this belief. We see differences between the real and artificial maps which are both qualitative and quantitative. 

First of all, for $\eps=0.08$ we find two directions around which there is a significant concentration of circular structures distinguished by the large values of $I_{(\eps,\ga_0)}(\theta,\phi)$ confirming the earlier findings in \cite{MNR}. The Galactic coordinates of these two directions are approximately $(\tilde\theta_1,\tilde\phi_1)=(2.60,3.70)$ (which correlates with the so called ``cold spot'')
and $(\tilde\theta_2,\tilde\phi_2)=(2.59,2.89)$ (this spot with opposite values of the integral was first noticed in \cite{MNR} and is confirmed by the present analysis). Actually, looking at Figures 1 and 2, one sees two significant centers of the red circles. The center of the dominant red circles has coordinates $(\tilde\theta_2,\tilde\phi_2)=(2.59,2.89)$, and the center of the fainter ones is  $(\tilde\theta_3,\tilde\phi_3)=(2.40,3.08)$. 

\begin{figure}[htb]
\begin{center}
\includegraphics[scale=0.22]{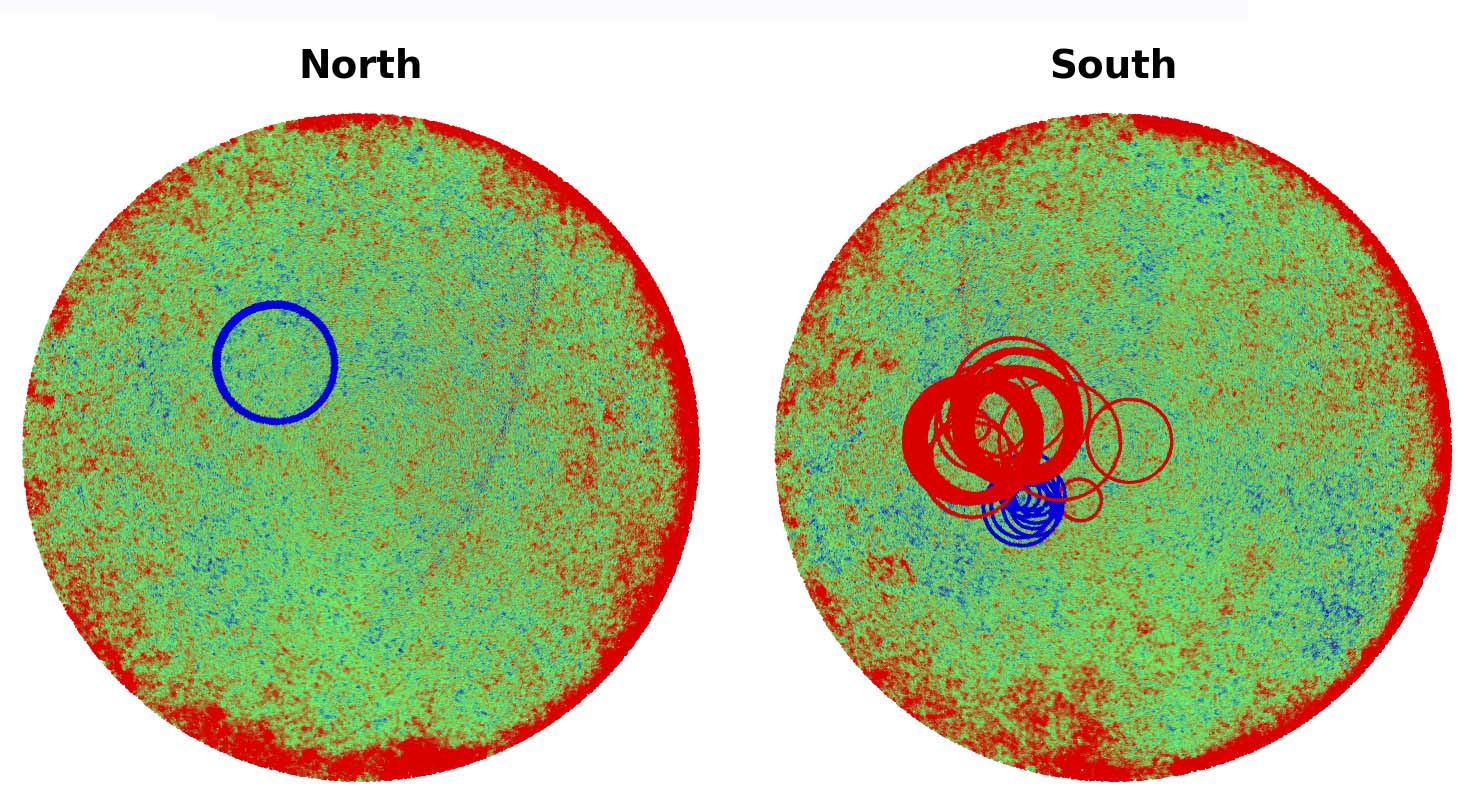}
\caption{\scriptsize{This is a Figure 1 in a different projection. We plotted rings having the width $\epsilon=0.08$ radians, as visible on the Planck CMB temperature maps in the frequency band 70GHz, but now in the stereographic projections from the respective Galactic Poles.}}
\end{center}
\end{figure}

The quantitative comparison between the real and simulated maps is deferred to the future publication \cite{AMNP}. Here we only note that for $\eps=0.08$ the measure $\si$ of the extremality of (positive or negative) values, as introduced in \cite{KM,MNR}, gives on the real map a value bigger than any $\si$ calculated on 298 (out of 300) artificial maps.
\begin{figure}[htb]
\begin{center}
\includegraphics[scale=0.22]{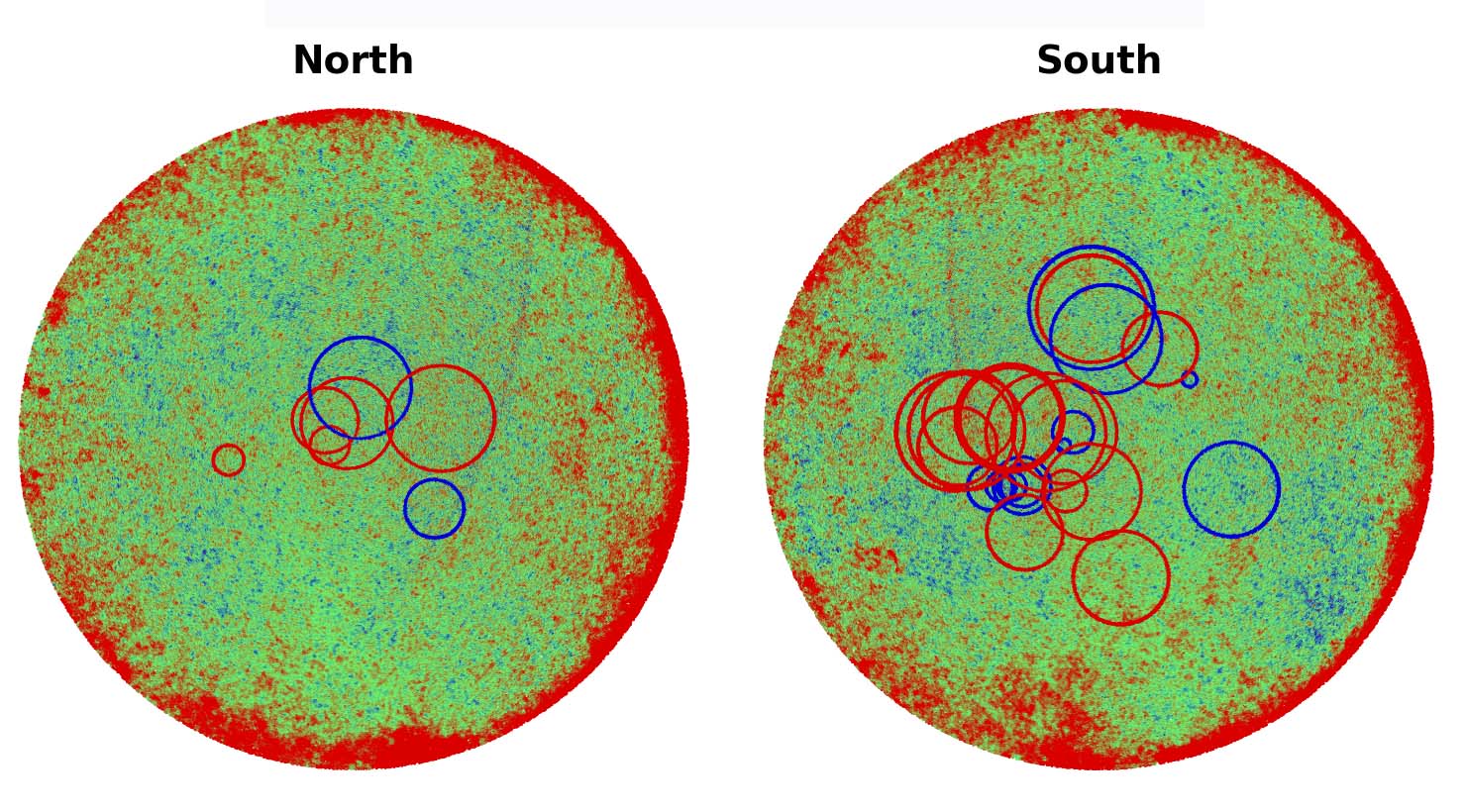}
\caption{\scriptsize{Plot of rings having the width $\epsilon=0.04$ as visible on the Planck temperature map with frequency band 70GHz, in the same stereographic projections as in Figure 2. Note that these rings are more scattered than those with the width $\epsilon=0.08$ radians. But observe that the rings which were present at the picture with $\epsilon=0.08$ have also their counterparts in the rings with this value of $\epsilon$.}}
\end{center}
\end{figure}

We can also note that the Southern Galactic hemisphere is remarkably different than the Northern one: the number of circular structures with large absolute values of the overlap integral is significantly larger in the South than in the North with respect to the Milky Way disk.

\begin{figure}[htb]
\begin{center}
\includegraphics[scale=0.22]{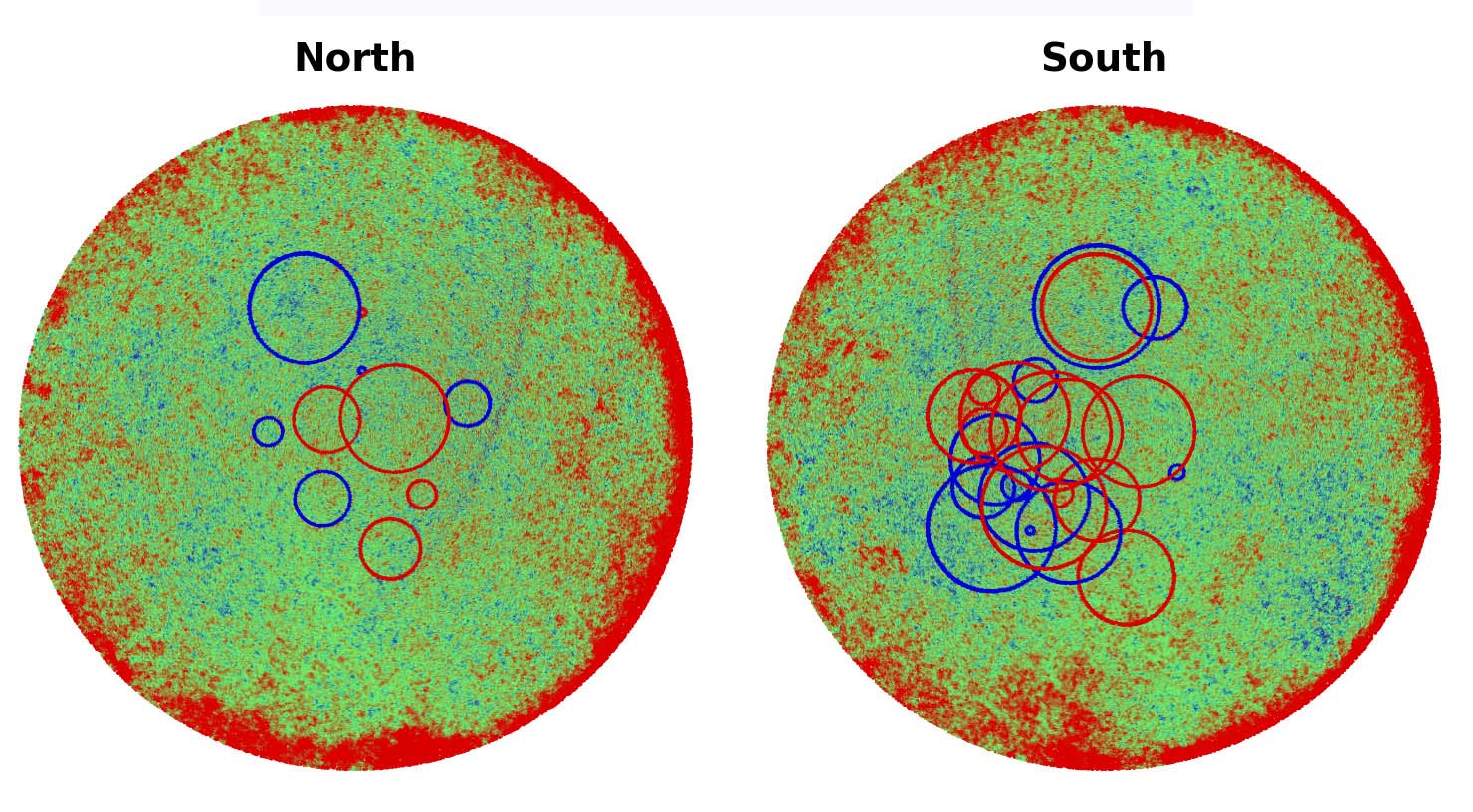}
\caption{\scriptsize{Plot of rings having the width $\epsilon=0.02$ as visible on the Planck map temperature with frequency band 70GHz, in the same stereographic projections as in Figure 2. Similar comment as in Figure 3 applies.}}
\end{center}
\end{figure}

In Figure 1 we have plotted the views of the circular structures on the Planck map with two projection with respect to the Milky Way disk.
In Figures 2, 3, 4 and 5 we have plotted the circular structures on the Planck map on the Southern and Northern hemispheres for $\eps=0.08,\ 0.04,\ 0.02,\ 0.01$ respectively (using the stereographic projection from the appropriate pole and inverting)

\begin{figure}[htb]
\begin{center}
\includegraphics[scale=0.30]{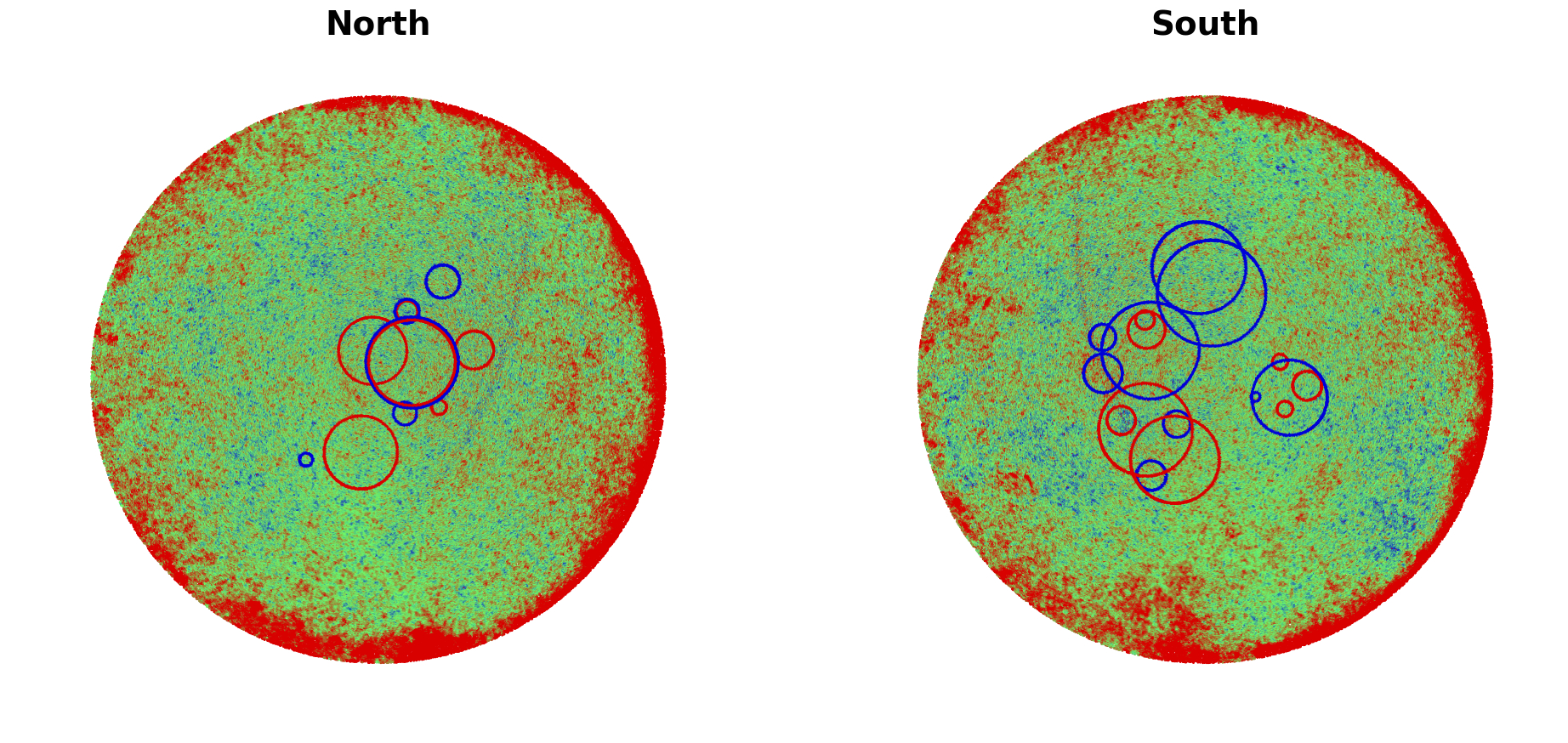}
\caption{\scriptsize{Plot of rings having the width $\epsilon=0.01$ as visible on the Planck map temperature with frequency band 70GHz, in the same stereographic projections as in Figure 2.}}
\end{center}
\end{figure}

\vspace{0.2cm}
\noindent{\bf {Acknowledgments:}} We gratefully acknowledge helpful discussions with Pawe{\l} Bielewicz, Marek Demia\'nski, C. Denson Hill, George Efstathiou, E. Ted Newman, Roger Penrose and  B{\l}a{\.z}ej Ruszczycki.


\begin{thebibliography}{99}

\bibitem{PC} Planck Collaboration, {\it Planck 2013 results}, http://planck.caltech.edu/publications2013Results.html

\bibitem{MNR} K. A. Meissner, P. Nurowski and B. Ruszczycki (2013) {\it Structures in the
microwave background radiation}, Proc. R. Soc. {\bf A469}:2155, 20130116 (2013),
arXiv:1207.2498[astro-ph.CO]

\bibitem{P} R. Penrose, {\it Cycles of Time: An Extraordinary New View of the Universe}, Bodley Head, 2010.

\bibitem{HE} K. M. G\'orski at al., {\it HEALPix: A Framework for High-Resolution Discretization and Fast Analysis of Data Distributed on the Sphere}, Astroph. J. {\bf 622} (2005) 759-771.

\bibitem{GP} V.G. Gurzadyan and R. Penrose (2013) {\it On CCC-predicted
concentric low-variance circles in CMB sky.} Eur. Phys. J. Plus (2013) 128, 22-38.

\bibitem{AMNP} D. An, K.A. Meissner, P. Nurowski and R. Penrose, {\it in preparation}

\bibitem{KM} K.A. Meissner, (2012) {\it A Tail Sensitive Test for Cumulative Distribution Functions}, http://arxiv.org/pdf/1206.4000.pdf
\end{thebibliography}
\end{document}